\documentclass[letterpaper, 10 pt, conference]{ieeeconf}  
\IEEEoverridecommandlockouts                             
\IEEEoverridecommandlockouts                             
\usepackage{graphics} 
\usepackage{epsfig} 
\usepackage{soul}
\usepackage{xcolor}
\usepackage{amsmath,amssymb,amsfonts}
\usepackage{multirow}

\def\BibTeX{{\rm B\kern-.05em{\sc i\kern-.025em b}\kern-.08em
    T\kern-.1667em\lower.7ex\hbox{E}\kern-.125emX}}

\def\BibTeX{{\rm B\kern-.05em{\sc i\kern-.025em b}\kern-.08em
    T\kern-.1667em\lower.7ex\hbox{E}\kern-.125emX}}

\begin{document}

\title{\LARGE \bf
Adaptive Spike-Like Representation of EEG Signals for \\Sleep Stages Scoring
}
\author{
Lingwei Zhu*, Koki Odani*, Ziwei Yang*, Guang Shi, Yirong Kan, Zheng Chen, and Renyuan Zhang
\thanks{This work was supported by KAKENHI (21K11809), Japan.}
\thanks{Lingwei Zhu, Koki Odani, Ziwei Yang, Guang Shi, Yirong Kan, Zheng Chen, and Renyuan Zhang are with Graduate School of Science and Technology, Nara Insitute of Science and Technology, Takayamacho 8916-5, Ikoma, 6300192 Japan.
(e-mail: chen.zheng.bn1@is.naist.jp)}%
\thanks{*Joint first authors.}
}

\maketitle

\begin{abstract}
Recently there has seen promising results on automatic stage scoring by extracting spatio-temporal features from electroencephalogram (EEG). Such methods entail laborious manual feature engineering and domain knowledge. In this study, we propose an adaptive scheme to probabilistically encode, filter and accumulate the input signals and weight the resultant features by the half-Gaussian probabilities of signal intensities. The adaptive representations are subsequently fed into a transformer model to automatically mine the relevance between features and corresponding stages. Extensive experiments on the largest public dataset against state-of-the-art methods validate the effectiveness of our proposed method and reveal promising future directions.
\end{abstract}

\begin{figure*}[t]
\centering
\includegraphics[width=0.90\linewidth]{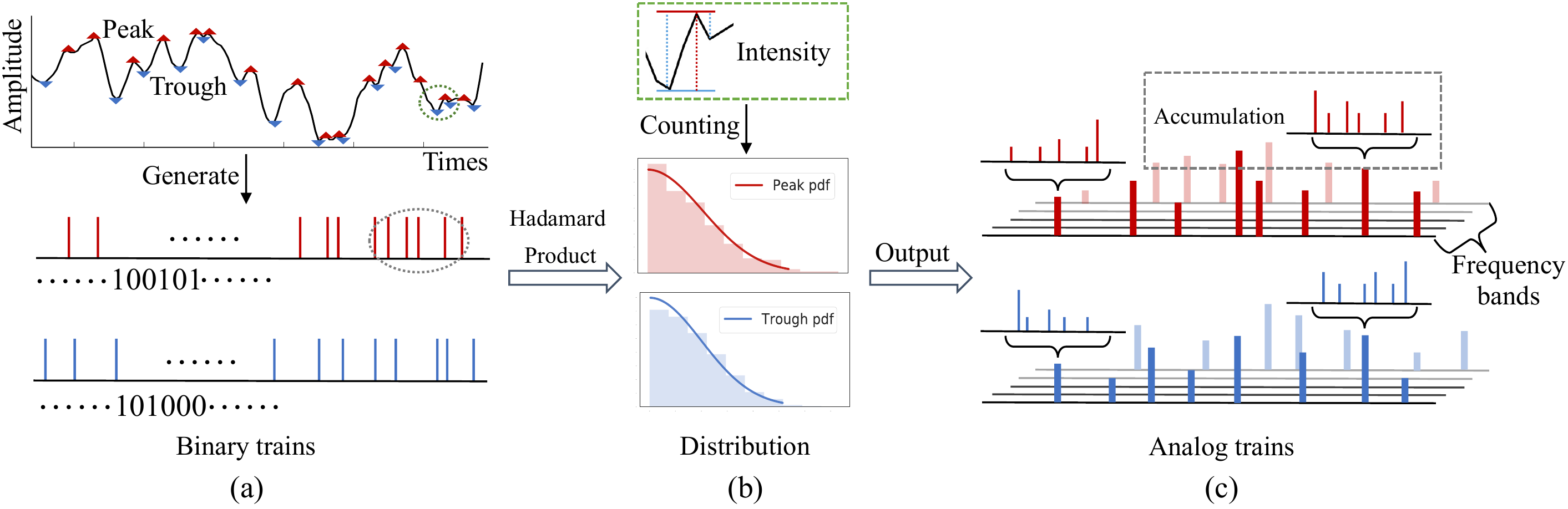}
\caption{The workflow converts EEG signals into novel peak trains representation.
pdf denotes the probability density function of the half-Gaussian distribution.
(a), (b) and (c) correspond to the Encode, Probabilitize and Accumulate phases described in Section \ref{sec:adaptive}, respectively.
}
\label{fig1}
\end{figure*}

\section{Introduction}
Sleep screening is not only an important tool in the assessment of pathophysiology but also an indispensable ingredient for the exploration of neuroscience \cite{Me2}.
Currently, the commonly accepted criterion for screening sleep is to decompose it into five different stages, i.e., wake, rapid eye movement (REM), or non-REM, where the non-REM stage can be further divided into N1, N2, and N3 according to the American Academy of Sleep Medicine (AASM) \cite{ct2}.
However, stage scoring typically requires visual inspection of stage-dependent characteristics, especially in electroencephalogram (EEG), and manual scoring from experts on a 30-second-epoch basis.
This time-consuming manual process is a major obstacle for advancing our understanding of sleep and more importantly, for deploying sleep-scientific findings into neuroscientific and pathological problems \cite{apnea}.
 
Recently the sleep community has seen promising results of automatic stage scoring/classification by extracting spatio-temporal features from EEG recordings \cite{sp3}.
However, those methods have not been widely deployed to the clinical setting due to their failure to consistently deliver a satisfying yet interpretable outcome.
In fact, the reason for such unsatisfying results has been hinted at by \cite{snndelayneuron}.
There is a missing piece of description for sleep stage-dependent characteristics that is usually ignored:
the internally generated fluctuations of neuronal excitability, which is on top of temporal and spatial patterns and difficult to capture. 
Though temporal and spatial patterns along are informative, these fluctuations are important themselves in providing  complementary information about the \emph{dynamics} in brain neurons.

Capturing the dynamic fluctuations is non-trivial since sleep stage-dependent patterns vary greatly along the spatio-temporal axes \cite{wake}.
Such patterns appear in random positions with random duration: the unpredictability of such features even motivated the introduction of chaotic theory for describing their behavior \cite{TKDE}.
As a concrete example, consider K-complex (with duration 0.5$\sim$1.5 seconds)  whose appearance signals the corresponding epoch should be scored as stage N2.
However, their appearance are temporally random and transient, hence great variability exists in the EEG recordings \cite{trasientissue}.
Conventional methods inductively construct feature vectors by statistical procedures \cite{psd2specenstas}, or by automatic feature extraction via deep neural networks \cite{sp1}.
However, such methods are all \emph{static} in that they do not attempt to capture the dynamic fluctuations.

In this paper we propose to capture the dynamic fluctutations in the EEG recordings by exploiting the \emph{spike-train} method \cite{TBME2021dynamic} which has been well-studied in other pathological literature, e.g., identifying seizure \cite{spikeseizure}.
This is due to its peak-and-trough feature extraction can reflect the neurodynamic activity underlying EEG data to some extent.
Specifically, we propose to probabilistically encode the intensity of peaks and troughs of EEG data by leveraging half-Gaussian distributions and feed the encoded features into a subsequent novel model for finding stage-dependent correlations in the spike-train sequence.
Extensive experiments on the largest public sleep dataset yields results competitive to the state-of-the-art methods, which validates our assumption that dynamic fluctuations are important and complementary to the existing spatio-temporal representation. 
To the best of our knowledge, this is the first attempt of introducing the spike-train mechanism into the sleep stage scoring community.

\section{Material and Method}
\subsection{Dataset and Preprocessing}
We use  the Sleep Heart Health Study (SHHS) in this paper\footnote{www.sleepdata.org}.
The SHHS database consists of two rounds of at-home PSG recordings (SHHS Visit 1 and SHHS Visit 2).
Due to the unscored epochs and misaligned records, here, we used only the SHHS Visit 1 containing two channel EEG records (C4-A1 and C3-A2) from 5793 subjects sampled at 125 Hz.
The records of the SHHS were manually annotated into six classes (W, S1, S2, S3, S4, REM).
Noteworthily, we merged S3 and S4 stages into one deep-sleep class referring to the AASM standards.

The original signal passed through one 8th order band-pass filter within 0.5 to 30-35 Hz, followed by five filters with cutoff frequencies as:
the delta waves (0-4 Hz), theta waves (4-8 Hz), alpha waves (8-12 Hz), sigma waves (12-16 Hz) and beta waves (16-32 Hz).
Every signal hence generated five filtered signals for subsequent processing, each corresponding to a frequency-domain feature of a sleep stage.

\subsection{Adaptive EEG Spike-Train}\label{sec:adaptive}

In this study, the EEG signals are converted into a spike-train representation.
The whole processing workflow is illustrated in Fig. \ref{fig1} and detailed below.

\textbf{Encode. } 
In order to find peaks and troughs in a sequence of $n$ amplitude measurements $x:=(x_1, \dots, x_n)$, we differentiate them to find zeros $ dx_{i} \!=\! 0$ and record their indices $i$.
Note that we have five such sequences.
We form a new sequence $y$ by inspecting the values of $dx_{i}$, let $y_i=1$ if $dx_{i}=0$ and 0 otherwise.
Hence, $y:=(\dots, 0, 1, 0, \dots) \in \mathbb{R}^n$, where $1$s are located at indices where $dx_{i}=0$.
This part corresponds to the workflow shown in Fig. \ref{fig1}(a).
The original intensity sequence $x$ and the binary mask $y$ are passed to the next layer of processing, i.e. subsequent probabilitizing.

\textbf{Probabilitize. } 
Given intensity $x$ and mask $y$, we propose to mask $x$ as $x \odot y=(\dots, 0, x_{i}, 0, \dots)\in\mathbb{R}^{n}$, where $\odot$ denotes element-wise product.
The masked intensities are mapped to values between $[0,1]$ by the half-Gaussian distribution:
\begin{align}
    f(z; \mu, \sigma) := \frac{\sqrt{2}}{\sigma\sqrt{\pi}} \exp\left(-\frac{z^2}{2\sigma^2}\right), \quad z\geq 0,
    \label{eq:half_gaussian}
\end{align}
where we let the input $z\!=\!x\odot y$ and write the probabilities as $f$ in short.
Eq. (\ref{eq:half_gaussian}) can be seen as a probabilistic encoding scheme in that we weight the contribution of input spike-train $z$ via probabilities, i.e. $z' \!:=\!f\odot z$.  
The flow of probabilitizing matches Fig. \ref{fig1}(b).
Intuitively, it quantitatively measures the morphological features of neuronal activation: consider a sequence of spikes of insignificant amplitude within a certain period, small amplitude implies small $f$ values, which leads to low accumulation.

\textbf{Accumulate. }
Since sleep stage-dependent waves are transient such as the K-complex introduced in Introduction, we leverage a short-duration non-overlapping sliding window to sequentially extract features from the weighted input spike-train $z'=(z'_{1}, \dots, z'_{n})$. 
Specifically, we use the sliding window of 25 steps as 
\begin{align}
    z' &= (\underbrace{z'_{1}, \dots, z'_{25}}_{a_1 = \sum_{j=1}^{25}z'_{j}}, \overbrace{ z'_{26} \dots, z'_{50}}^{a_{2}}, \dots, z'_{n}) 
\end{align}
where the resultant accumulated vector is denoted as $a\in\mathbb{R}^{\frac{n}{25}}$, and we assume the length $n$ is a multiple of 25.
The above-mentioned processing is respectively performed for peaks and troughs, and hence 10 sequences (five sequences each with its peaks and troughs) are generated in total. 
This accumulation part is illustrated in Fig. \ref{fig1}(c).

\begin{figure}[t]
	\centering
	\includegraphics[width=0.99\columnwidth]{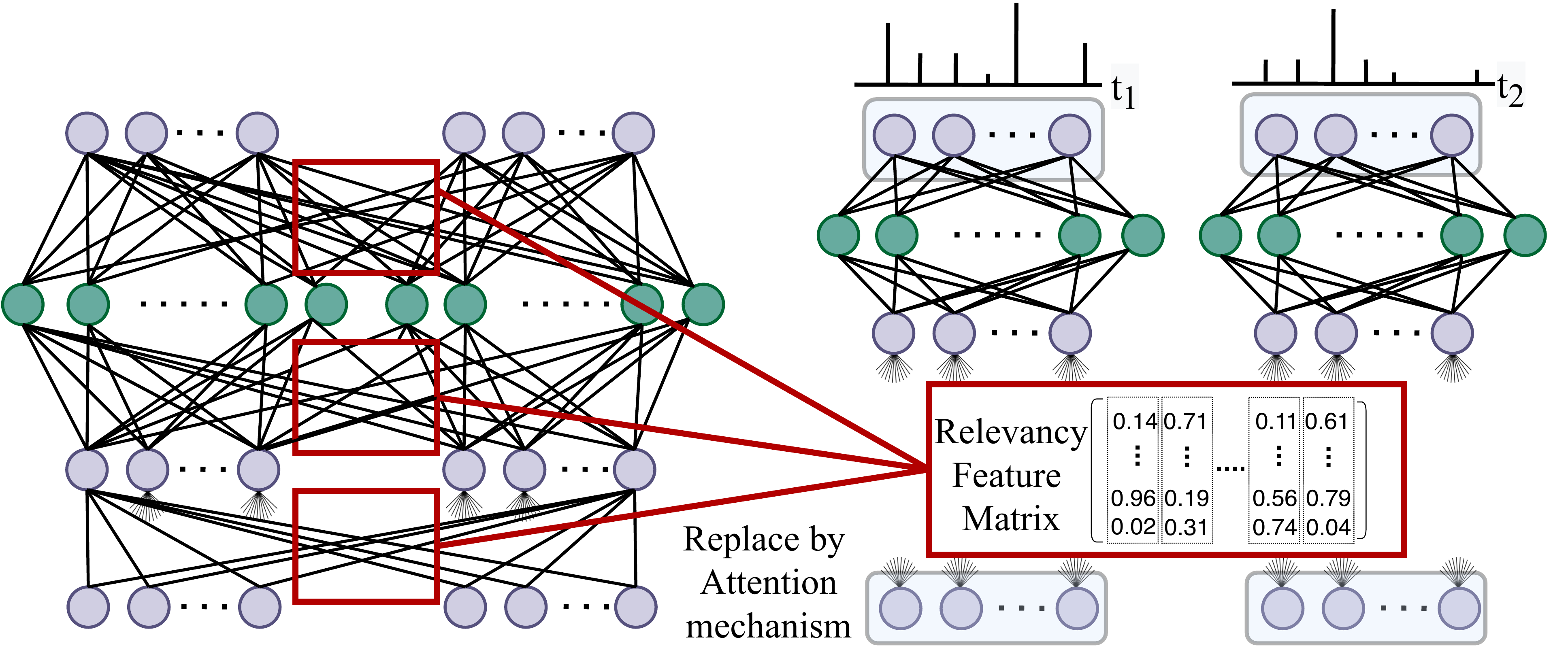}
	\caption{Architecture of the attention model}
	\label{fig2}
\end{figure} 

\begin{table*}
\centering
\caption{Performance obtained by the proposed model, and existing works using the same SHHS database. We make the best stage-wise performance of each evalution metric.}
\label{tb1}
\begin{tabular}{ccclcccccc} 
\hline\hline
\begin{tabular}[c]{@{}c@{}}\end{tabular} & Signal resource                                                                                 & \begin{tabular}[c]{@{}c@{}}Num. of \\subjects\end{tabular} &     & Wake          & N1            & N2            & N3            & REM            & \begin{tabular}[c]{@{}c@{}}Overall\\Accuracy\end{tabular}  \\ 
\hline
                                                 & \multirow{3}{*}{\begin{tabular}[c]{@{}c@{}}~EEG\end{tabular}}  & \multirow{3}{*}{140}                                       & \textit{Pre} & 0.89          & 0.55          & 0.75          & 0.84          & 0.78           & \multirow{3}{*}{0.83}                                      \\
Foroozan Karimazadeh et al,2017~\cite{Foroozan}                 &                                                                                        &                                                            & \textit{Re}  & 0.81          & 0.58          & 0.68          & 0.54          & 0.78           &                                                            \\
                                                 &                                                                                        &                                                            & \textit{F1}  & 0.85          & 0.56          & 0.71          & 0.66          & 0.78           &                                                            \\ 
\hline
                                                 & \multirow{3}{*}{\begin{tabular}[c]{@{}c@{}}EEG, EOG,\\EMG\end{tabular}}                 & \multirow{3}{*}{15804}                                     & \textit{Pre} & 0.90          & \textbf{0.69} & 0.84          & 0.80          & 0.79           & \multirow{3}{*}{\textbf{0.86}}                             \\
Siddharth Biswal et al,2018~\cite{Siddharth}                     &                                                                                        &                                                            & Re  & 0.81          & \textbf{0.67} & 0.78          & 0.76          & 0.74           &                                                            \\
                                                 &                                                                                        &                                                            & \textit{F1}  & 0.85          & \textbf{0.68} & 0.81          & 0.78          & 0.76           &                                                            \\ 
\hline
                                                 & \multirow{3}{*}{\begin{tabular}[c]{@{}c@{}}EEG\end{tabular}}                 & \multirow{3}{*}{5804}                                      & \textit{Pre} & 0.90          & 0.54          & 0.84          & 0.83          & 0.77           & \multirow{3}{*}{0.84}                                      \\
Enrique et al,2020~\cite{Enrique}                              &                                                                                        &                                                            & \textit{Re}  & 0.91          & 0.22          & \textbf{0.91} & 0.65          & 0.88           &                                                            \\
                                                 &                                                                                        &                                                            & \textit{F1}  & 0.91          & 0.31          & \textbf{0.87} & 0.73          & 0.81           &                                                            \\ 
\hline
                                                 & \multirow{3}{*}{\begin{tabular}[c]{@{}c@{}}EEG\end{tabular}}                 & \multirow{3}{*}{329}                                       & \textit{Pre} & 0.90          & 0.30          & 0.85          & 0.85          & \textbf{0.80}  & \multirow{3}{*}{0.85}                                      \\
Emadeldeen Eldele et al,2021~\cite{Emadeldeen}                    &                                                                                        &                                                            & \textit{Re}  & 0.83          & 0.36          & 0.86          & 0.85          & \textbf{0.81} &                                                            \\
                                                 &                                                                                        &                                                            & \textit{F1}  & 0.86          & 0.33          & 0.85          & 0.85          & \textbf{0.80 } &                                                            \\ 
\hline
                                                 & \multirow{3}{*}{\begin{tabular}[c]{@{}c@{}}EEG\end{tabular}} & \multirow{3}{*}{5736}                                      & \textit{Pre} & \textbf{0.91} & 0.31          & \textbf{0.86} & \textbf{0.85} & 0.79           & \multirow{3}{*}{0.85}                                      \\
This study                                &                                                                                        &                                                            & \textit{Re}  & \textbf{0.92} & 0.35          & 0.80          & \textbf{0.87} & 0.78           &                                                            \\
                                                 &                                                                                        &                                                            & \textit{F1}  & \textbf{0.92} & 0.33          & 0.83          & \textbf{0.86} & 0.78           &                                                            \\
\hline\hline
\end{tabular}
\end{table*}

\subsection{Transformer Model}
To parallel capture the trasient stage-dependent features in adaptive spike-train, we apply the popular attention-based model in our automatic staging system \cite{VIT}.

The model architecture has two sub-layers: a multi-head self-attention layer and a feed-forward layer.
Here, the self-attention mechanism is implemented by three matrices: query ($Q$), key ($K$), and value ($V$).
The attention function utilizes each row of $Q$ to query $K$ in parallel to collect the correlations among the patches.
A set of the weights computed by the softmax of ($Q, K$) pairs are assigned to $V$.
Finally, the weighted value matrix encompasses the importance of embeddings that can be fed into next layer.
The modeling of these patches, called as scaled dot-product attention, is given by Eq. \eqref{eq2} by \cite{VIT}:
\begin{equation}
\texttt{Attention}(Q, K, V) = \texttt{softmax}(\frac{QK^{T}}{\sqrt{d}})V\label{eq2}
\end{equation}
where $d$ denotes a scaling coefficient that is used to mitigate the gradient vanishing problem of the dot products of query with key. 
Afterwards, the resultant attention matrix is fed into a subsequent feed-forward layer.
As shown in Fig \ref{fig2}, this layer is column-wise a fully connected network that yields the a set of parameterized weights to fit the relevancy function between one short time duration and others.
Meanwhile, the subsequent non-linear activation function is GELU.



\section{Experiment and Result}
\subsection{Experimental Setup}
The hyperparameters used for experiments are listed in Table \ref{tb2}.
We implemented a record-wise 7-fold cross-validation. 
For each trial, we used a test set consisting of 800 subjects. 
Three metrics were used to evaluate the performance: the stage-specific precision (Pre), recall (Re), and macro-averaging F1-score (F1).
To investigate the effect of the half-Gaussian filters, we constructed a control group where the spike-train was generated with a fixed cut-off threshold.
We also compared the model performance against the state-of-art works that adopted the same experiment database (i.e., SHHS).

\begin{table}
\centering
\caption{Parameters description}
\label{tb2}
\begin{tabular}{c|l|c} 
\hline\hline
\multicolumn{1}{l|}{}                  & \textbf{Parameter}                    & \textbf{Value}  \\ 
\hline
\multirow{4}{*}{Spike-train} & Half-Gaussian filter mean             & 0               \\
                                       & Half-Gaussian filter variance         & 0.5             \\
                                       & Half-Gaussian filter window size      & 125             \\
                                       & Accumulation width            & 25              \\ 
\hline
\multirow{6}{*}{Attention}           & \#Stacked encoder                     & 8               \\
                                       & \#Heads ($h$)                         & 4               \\
                                       & Dimension of linear projection of $D$ & 128              \\
                                       & Normalization-like scale ($\sqrt{d}$) & 8               \\
                                       & Dimension of MLP output               & 128             \\
                                       & Dropout rate                          & 0.5             \\ 
\hline
\multirow{3}{*}{Training}        & \#Epoch                      & 100             \\
                                       & Batch size                            & 32              \\
                                       & Learning rate                         & $10^{-4}$         \\
\hline\hline
\end{tabular}
\end{table}

\begin{figure}[t]
	\centering
	\includegraphics[width=.98\linewidth]{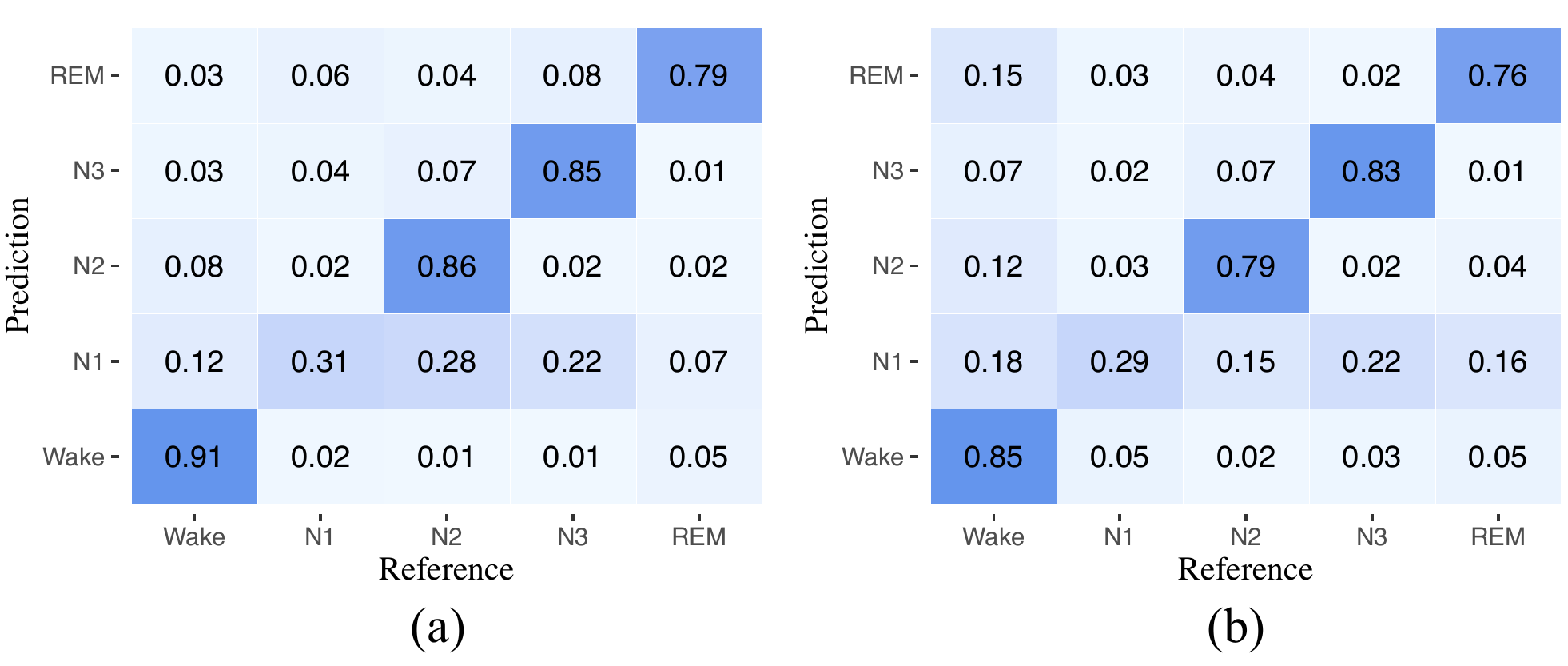}
	\caption{(a) The confusion matrix resulted from the proposed model where the peak train was generated through the half-Gaussian filter. (b) The confusion matrix resulted from the proposed model but the peak train was generated by a fixed cut-off threshold.}
	\label{fig3}
\end{figure} 

\begin{figure*}[t]
	\centering
	\includegraphics[width=0.85\textwidth]{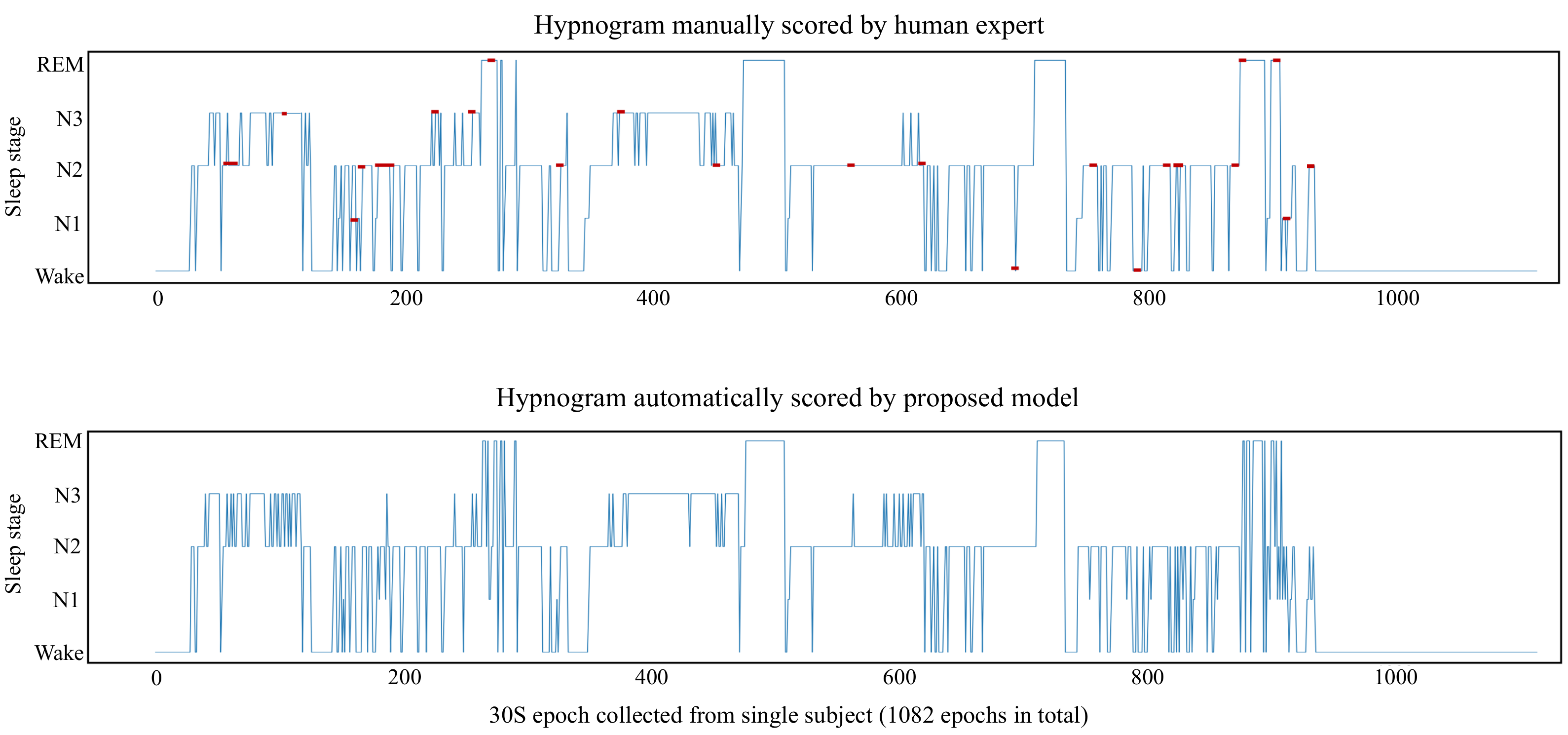}
	\caption{Examples of the hypnogram manually scored by human expert (top) and the hypnogram automatically scored by the proposed model (bottom) for a single subject from the SHHS dataset. Misclassification is marked in red. The accuracy and F1-score of our model are 0.86 and 0.85, respectively.
	}
	\label{fig4}
\end{figure*} 

\subsection{Result}
Fig. \ref{fig3}(a) shows the confusion matrix output by the proposed model.
Compared with the ablation in Fig. \ref{fig3}(b) that generated spike-train with a fixed cut-off threshold,
we find that removing the half-Gaussian filter affects the classification of all five sleep stages, which suggests the half-Gaussian filter played a crucial role in generating more reliable adaptive spike-train features, testifying to that the probabilistic encoding aids in extracting dynamic fluctuations.

Table \ref{tb1} shows the performance statistics of the proposed model. 
The results show that our model achieved better performance against other combinations of popular feature extraction methods and sequential models. 
We observe our classification results on the stage Wake established a new SOTA compared with other baseline methods.
A closer look on other stages indicates the proposed model achieved competitive results on stage N2 and REM with the F1-score being 0.83 and 0.78.

Fig. \ref{fig4} shows a visualization of hypnograms examples scored by a human expert and our classifier, for a subject from the SHHS database.
We observe that most of the misclassified stages are between stages N2 and other stages (N3 and REM).
Moreover, misclassification often occurred when the sleep stages transferred frequently in a relatively short period.
In contrast, our model is robust for relative steady micro-states (non-REM to REM) before stage transition.

\section{Conclusion}
Developing automatic sleep staging methods is important for both pathological diagnosis and neuroscientific research.
Though recent work achieved promising results with deep learning, the performance relied crucially on manually selected spatio-temporal features.
We proposed a novel method of adaptive spike-train for extracting dynamic fluctuations which were considered complementary to existing EEG representations. 
Extensive experiments validated the effectiveness of the proposal on the largest public sleep dataset against state-of-the-art methods.

One promising future direction is to incorporate between-epoch information such as the sequential order of the stages, e.g. REM cannot transfer to N1. 
Augmentation with such information might potentially improve the performance.

\bibliographystyle{IEEEtran}
\bibliography{IEEEabrv,Bibliography}

\end{document}